\documentclass[preprint2]{aastex62}

\begin{document}

\title{Self-similarity and power-laws in GRB 190114C}

\author{R. Ruffini}
\affiliation{ICRANet, P.zza della Repubblica 10, 65122 Pescara, Italy}
\affiliation{ICRA and Dipartimento di Fisica, Sapienza Universit\`a di Roma, P.le Aldo Moro 5, 00185 Rome, Italy}
\affiliation{Universit\'e de Nice Sophia Antipolis, CEDEX 2, Grand Ch\^{a}teau Parc Valrose, Nice, France}
\affiliation{ICRANet-Rio, Centro Brasileiro de Pesquisas F\'isicas, Rua Dr. Xavier Sigaud 150, 22290--180 Rio de Janeiro, Brazil}
\affiliation{INAF, Viale del Parco Mellini 84, 00136 Rome, Italy.}

\author{Liang Li}
\affiliation{ICRANet, P.zza della Repubblica 10, 65122 Pescara, Italy}

\author{R. Moradi}
\affiliation{ICRANet, P.zza della Repubblica 10, 65122 Pescara, Italy}
\affiliation{ICRA and Dipartimento di Fisica, Sapienza Universit\`a di Roma, P.le Aldo Moro 5, 00185 Rome, Italy}
\affiliation{INAF -- Osservatorio Astronomico d'Abruzzo,
Via M. Maggini snc, I-64100, Teramo, Italy}

\author{J.~A.~Rueda}
\affiliation{ICRANet, P.zza della Repubblica 10, 65122 Pescara, Italy}
\affiliation{ICRA and Dipartimento di Fisica, Sapienza Universit\`a di Roma, P.le Aldo Moro 5, 00185 Rome, Italy}
\affiliation{ICRANet-Rio, Centro Brasileiro de Pesquisas F\'isicas, Rua Dr. Xavier Sigaud 150, 22290--180 Rio de Janeiro, Brazil}
\affiliation{INAF -- Istituto di Astrofisica e Planetologia Spaziali, 00133 Via del Fosso del Cavaliere, 100, Rome, Italy}

\author{Yu~Wang}
\affiliation{ICRANet, P.zza della Repubblica 10, 65122 Pescara, Italy}
\affiliation{ICRA and Dipartimento di Fisica, Sapienza Universit\`a di Roma, P.le Aldo Moro 5, 00185 Rome, Italy}
\affiliation{INAF -- Osservatorio Astronomico d'Abruzzo,
Via M. Maggini snc, I-64100, Teramo, Italy}

\author{S.~S.~Xue}
\affiliation{ICRANet, P.zza della Repubblica 10, 65122 Pescara, Italy}
\affiliation{ICRA and Dipartimento di Fisica, Sapienza Universit\`a di Roma, P.le Aldo Moro 5, 00185 Rome, Italy}
\affiliation{INAF -- Istituto di Astrofisica e Planetologia Spaziali, 00133 Via del Fosso del Cavaliere, 100, Rome, Italy}

\author{ C.~L.~Bianco}
\affiliation{ICRANet, P.zza della Repubblica 10, 65122 Pescara, Italy}
\affiliation{ICRA and Dipartimento di Fisica, Sapienza Universit\`a di Roma, P.le Aldo Moro 5, 00185 Rome, Italy}
\affiliation{INAF -- Istituto di Astrofisica e Planetologia Spaziali, 00133 Via del Fosso del Cavaliere, 100, Rome, Italy}

\author{S.~Campion}
\affiliation{ICRANet, P.zza della Repubblica 10, 65122 Pescara, Italy}
\affiliation{ICRA and Dipartimento di Fisica, Sapienza Universit\`a di Roma, P.le Aldo Moro 5, 00185 Rome, Italy}

\author{J. D. Melon Fuksman}
\affiliation{ICRANet, P.zza della Repubblica 10, 65122 Pescara, Italy}
\affiliation{ICRA and Dipartimento di Fisica, Sapienza Universit\`a di Roma, P.le Aldo Moro 5, 00185 Rome, Italy}

\author{C.~Cherubini}
\author{S.~Filippi}
\affiliation{ICRANet, P.zza della Repubblica 10, 65122 Pescara, Italy}
\affiliation{ICRA and Department of Engineering, University Campus Bio-Medico of Rome, Via Alvaro del Portillo 21, 00128 Rome, Italy}

\author{M.~Karlica}
\affiliation{ICRANet, P.zza della Repubblica 10, 65122 Pescara, Italy}
\affiliation{ICRA and Dipartimento di Fisica, Sapienza Universit\`a di Roma, P.le Aldo Moro 5, 00185 Rome, Italy}
\affiliation{Universit\'e de Nice Sophia Antipolis, CEDEX 2, Grand Ch\^{a}teau Parc Valrose, Nice, France}

\author{ N.~Sahakyan}
\affiliation{ICRANet, P.zza della Repubblica 10, 65122 Pescara, Italy}
\affiliation{ICRANet-Armenia, Marshall Baghramian Avenue 24a, Yerevan 0019, Armenia}

\date{\today}

\begin{abstract}

Following Fermi and NOT observations, \citet{GCN23715} soon identified GRB 190114C as BdHN I at $z=0.424$. It has been observed since, with unprecedented accuracy, by Swift, Fermi-GBM and Fermi-LAT and  MAGIC in the MeV, GeV and TeV ranges all the way to the successful optical observation of our predicted supernova (SN). This GRB is a twin of GRB 130427A. Here we take advantage of the GBM data and  identify in it three different  ``Episodes''. Episode 1 represents the ``precursor'' which  includes the SN breakout and the creation of the new neutron star ($\nu$NS), the hypercritical accretion of the SN ejecta onto the NS binary companion, exceeding the NS  critical mass at $t_{\rm rf}=1.9$~s. Episode 2 starting at  $t_{\rm rf}=1.9$~s includes three major events: the formation of the BH, the onset of the  GeV emission and the onset of the ultra-relativistic prompt emission (UPE), which extends all the way up to $t_{\rm rf}=3.99$~s. Episode 3  which occurs at times following $t_{\rm rf}=3.99$~s reveals the presence of a ``cavity'' carved out in the SN ejecta by the BH formation. We perform an in depth time-resolved spectral analysis on the entire UPE with the corresponding determination of the spectra best fit by a cut-off power-law and a black body (CPL+BB) model,  and then we repeat the spectral analysis  in 5 successive time  iterations in increasingly shorter time bins: we find a similarity in the spectra in each stage of the iteration revealing clearly a  self-similar structure. We find a power-law dependence  of the BB temperature  with index $-1.56\pm 0.38$, a dependence with index $-1.20\pm 0.26$ for the gamma-ray luminosity confirming a similar dependence  with index $-1.20\pm 0.36$ which we find  as well in the GeV luminosity, both expressed  in the rest-frame. We thus discover in the realm of relativistic astrophysics the existence of a self-similar physical process and power-law dependencies, extensively described in the micro-physical world by the classical works of Heisenberg-Landau-Wilson.
\end{abstract}

\section{Introduction}

On 14 January 2019, GRB 190114C was announced by the Swift satellite team \citep{GCN23688}. Its distance (redshift $z=0.42$) was determined a few hours later by the Nordic Optical Telescope located at the Canary Islands, Spain \citep{GCN23695}. Soon after, \citet{GCN23715} recognized that this source was a BdHN I, and ICRANet sent the GCN anticipating the possibility of the appearance of an associated supernova. The supernova was indeed detected at exactly the predicted time, as reported by \citet{GCN23983} on 19 March 2019. 

Recall that all GRBs have been classified into eight different equivalence classes \citep{2016ApJ...832..136R,2019ApJ...874...39W}, the BdHN are: long GRBs with a progenitor composed of a CO$_{\rm core}$ of $\sim 10\,M_\odot$ undergoing a supernova (SN) explosion leading to the formation of a new neutron star($\nu$NS). The CO$_{\rm core}$ is in a tight binary system with a binary NS companion, the binary period being as short as $4$ minutes. In the BdHN I class, the hypercritical accretion of the SN ejecta onto the NS brings it to exceed the critical mass to form a BH and the associated isotropic energy ($E_{\rm iso}$) is always larger than $10^{52}$~erg. When the NS mass increases, but remains below the critical mass, a BdHN II with $E_{\rm iso}$ smaller than $10^{52}$~erg is formed \citep{2018ApJ...859...30R, 2019ApJ...874...39W}.

In both cases, the hypercritical accretion on the $\nu$NS gives origin to the afterglow emission \citep{2019ApJ...874...39W}. The introduction of successive Episodes each characterized by a specific physical process has allowed the comprehension of X-ray flares \citep{2018ApJ...852...53R}, of gamma-ray flares, as well as to the identification of the transition from a SN to an hypernova \citep{2018ApJ...869..151R}.
 
Another useful contribution has been pointing out the crucial dependence of the GRB description on its viewing angle: in the plane or orthogonal to the plane of the binary progenitors \citep{2018ApJ...852...53R, 2018ApJ...869..151R} \citep{2016ApJ...833..107B,2019ApJ...871...14B} and as a function of the rotational period of the binary system $\sim$ 300 s \citep{2018ApJ...869..151R}.

This progress has been made possible thanks also to the visualisation and simulation technique developed in our prolonged collaboration with LANL \citep{2014ApJ...793L..36F, 2015PhRvL.115w1102F, 2016ApJ...833..107B, 2018IJMPA..3344031R, 2018ARep...62..840B, 2019ApJ...871...14B} following in great detail the SN hypercritical accretion onto the companion NS in the BdHN \citep{2016ApJ...832..136R,2015ApJ...812..100B, 2018ApJ...859...30R}. 
 
In the case of GRB 190114C and GRB 130427A, these sources are seen ``from the top'' with a viewing angle normal to the plane of the orbit of the binary progenitors.

In the case of GRB 190114C, all phases of the BdHN I, starting from the onset of the SN breakout, to the accretion process, to the moment of formation of the BH, to the observations of gamma and  GeV emissions, to the  afterglow to the final identification of the optical SN have become observable with unprecedented  precision.

GRB160625B and GRB160509A, two BdHN I have been added in our analysis in order to prove the validity and generality of our approach applied to GRB190114C and GRB130427A. 

\begin{deluxetable*}{ccccccccccc}
\tabletypesize{\scriptsize}
\tablewidth{0pt}
\tablecaption{Parameters of the thermal shock breakout for the four precursors of the selected BdHNe I, details in the companion paper Li et al., (2019). In the first column the GRB name. In the second column and following ones the duration, respectively, in the rest frame and in the laboratory frame, the flux, the temperature, and the energy of the shock breakout $E_{sh}$. In the sixth column the GRB $E_{iso}$, well above the minimum energy of $10^{52}$ erg for the BdHN I and in the seventh column the redshift. In the last column, the evidence of the SN in GRB 190114C and GRB 130427A, directly derived from the optical observations. In the case of GRB 160509A and GRB 160625B the evidence of the SN is indirectly inferred from the mass and the spin of the $\nu$NS in the afterglows (see R. Ruffini, M. Karlica, et al. 2019, in preparation). Particularly significant is the proportionality of the $E_{sh}$ to the rotational energy of the co-rotating CO$_{\rm core}$ in the binary system of period $P\sim 4$~min. For the spectral analysis technique see caption of Table 2.}
\tablehead{
\colhead{GRB}
&\colhead{Duration}
&\colhead{Duration}
&\colhead{Flux}
&\colhead{Temperature}
&\colhead{$E_{\rm sh}$}
&\colhead{$E_{\rm iso}$}
&\colhead{redshift}
&\colhead{SN evidence}\\
&
\colhead{(s)}
&
\colhead{(s)}
&\colhead{(erg cm$^{-2}$ s$^{-1}$)}
&\colhead{(keV)}
&\colhead{(10$^{52}$ erg)}
&\colhead{(erg)}\\
&\colhead{(Rest)}
&\colhead{(Observation)}
&
&\colhead{(Rest)}
&\colhead{(Shock Breakout)}
&\colhead{(Total)}
&
&\colhead{}
}
\startdata
190114C & 1.12$\sim$1.68 & 0.39 &     1.06$^{+0.20}_{-0.20}$(10$^{-4}$)&27.4$^{+45.4}_{-25.6}$&2.82$^{+0.13}_{-0.13}$&(2.48$\pm$0.20)$\times$10$^{53}$&0.424&\cite{GCN23983}\\
\hline
130427A&0.0$\sim$2.6&1.94&2.14$^{+0.28}_{-0.26}$(10$^{-5}$)&44.91$^{+1.51}_{-1.51}$&0.65$^{+0.17}_{-0.17}$&$\sim$1.40$\times$10$^{54}$&0.3399&\cite{2013ApJ...776...98X}\\
160509A&0.0$\sim$7.0&3.23&7.9$^{+6.0}_{-3.1}$(10$^{-7}$)&...&2.96$^{+0.6}_{-0.6}$&$\sim$1.06$\times$10$^{54}$&1.17&Inferred from $\nu$NS\\
160625B&-1.0$\sim$2.0&1.25&6.8$^{+1.6}_{-1.6}$(10$^{-7}$)&36.8$^{+1.9}_{-1.9}$&1.09$^{+0.2}_{-0.2}$&$\sim$3.00$\times$10$^{54}$&1.406&Inferred from $\nu$NS\\
\enddata
\end{deluxetable*}

\section{GRB 190114C and GRB 130427A}\label{sec:190114C_130427A}

In our study of GRB 190114C we are facilitated by our previous analysis of GRB 130427A \citep{2018arXiv181200354R} which is a twin BdHN. While the Fermi-LAT observations of both of these BdHNe I are excellent, the Fermi-GBM data are more complete for GRB 190114C and allows the necessary detailed analysis to be performed.

In GRB 130427A with a very high isotropic energy of $E_{\rm iso}=1.4\times 10^{54}$~erg and $z=0.34$, the event count rate of n9 and n10 of Fermi-GBM in the radiation between $T_0 + 4.5$~s and $T_0 + 11.5$~s surpasses $\sim 8\times 10^4$ counts per second, where $T_0$ is the Fermi-GBM trigger time (see below). Consequently, the GBM data  were affected by pile up, which significantly deforms the spectrum in the aforementioned time interval \citep[see e.g. ][]{2014Sci...343...42A,2015ApJ...798...10R}. We were then in no condition to perform the detailed spectral analysis on that source necessary for identifying the underline physical origin of the MeV emission. 

On the other hand the GeV emission observed by LAT are excellent both for GRB 130427A and GRB 190114C. This allows us to adopt our previous analysis of the GeV radiation performed on GRB 130427A \citep{2015ApJ...798...10R, 2018arXiv181200354R, 2019ApJ...874...39W} for GRB190114C. 

For the case of GRB 190114C with $z=0.42$, we estimate an isotropic energy of $E_{\rm iso}=(2.48 \pm 0.22) \times 10^{53}$erg. More favorable observational conditions prevail here. The lower isotropic energy and the farther distance have converged to lead to  an optimal result with event count rate less than $\sim 3\times 10^4$ for the four brightest NaI detectors (n3, n4, n7 and n8) and one BGO detector (B0). We have performed the fully Bayesian analysis package, namely, the Multi-Mission Maximum Likelihood Framework (3ML, \citealt{2015arXiv150708343V})  as the main tool to carry out the temporal and spectral analyses for Fermi-GBM data \citep[e.g.,][]{2018arXiv181003129L, 2018arXiv181007313Y}. The GBM carries 12 sodium iodide (NaI, 8keV-1MeV) and 2 bismuth germinate (BG0, 200 keV-40 Mev) scintillation detectors \citep{2009ApJ...702..791M}. The four brightest triggered-NaI detectors (n3, n4, n7, n8) and one triggered-BGO detector (B0) are used to conduct the spectral analysis.
The background is selected, adopting the data before and after the burst, and fitted with a polynomial function with automatically determined order by 3ML. We select the source as the time interval of $T_{90}$ \citep{GCN23707}. The maximum likelihood-based statistics are used, the so-called Pgstat, given by a Poisson (observation)-Gaussian (background) profile likelihood \citep{1979ApJ...228..939C}. We show the light-curve of GBM data of GRB 190114C, time coordinate is converted to the rest-frame, see Fig.~\ref{thermal}.

\section{The 3 Episodes of GRB 190114C}\label{sec:3episodes}

\begin{figure*}
\centering
\includegraphics[angle=0, scale=0.5]{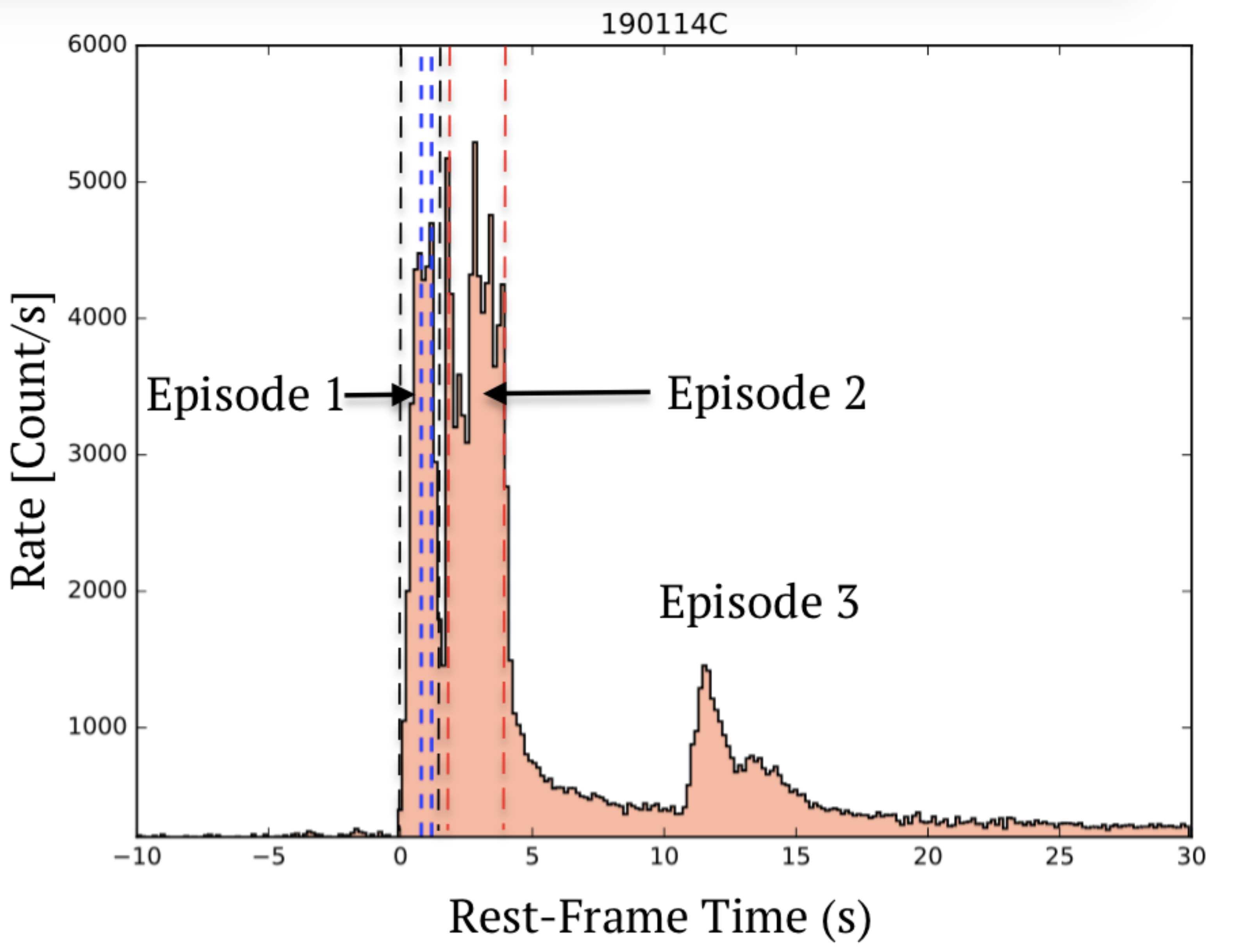}
\caption{The proposed three new episodes of GRB 190114C are here presented as a function of the rest-frame time. Episode 1 occurs $t_{\rm rf}=0$~s and $~t_{\rm rf}=1.9$~s, marked by dashed gray vertical lines. The blue dashed vertical lines represents the SN breakout. Episode 2 occurs from $t_{\rm rf}=1.9$~s to $~t_{\rm rf}=3.99$~s, and includes the UPE emission, marked by dashed red vertical lines. Episode 3 occurs at times after $~t_{\rm rf}=3.99$~s, staring at $~t_{\rm rf}=11$~s and ending at $~t_{\rm rf}=20$~s.}\label{thermal}
\end{figure*}

We propose here three new episodes with specific spectral features to identify the leading physical process manifested in the GBM data, see Fig.~\ref{thermal}. 
 
\begin{figure*}
\centering
\includegraphics[angle=0, scale=0.45]{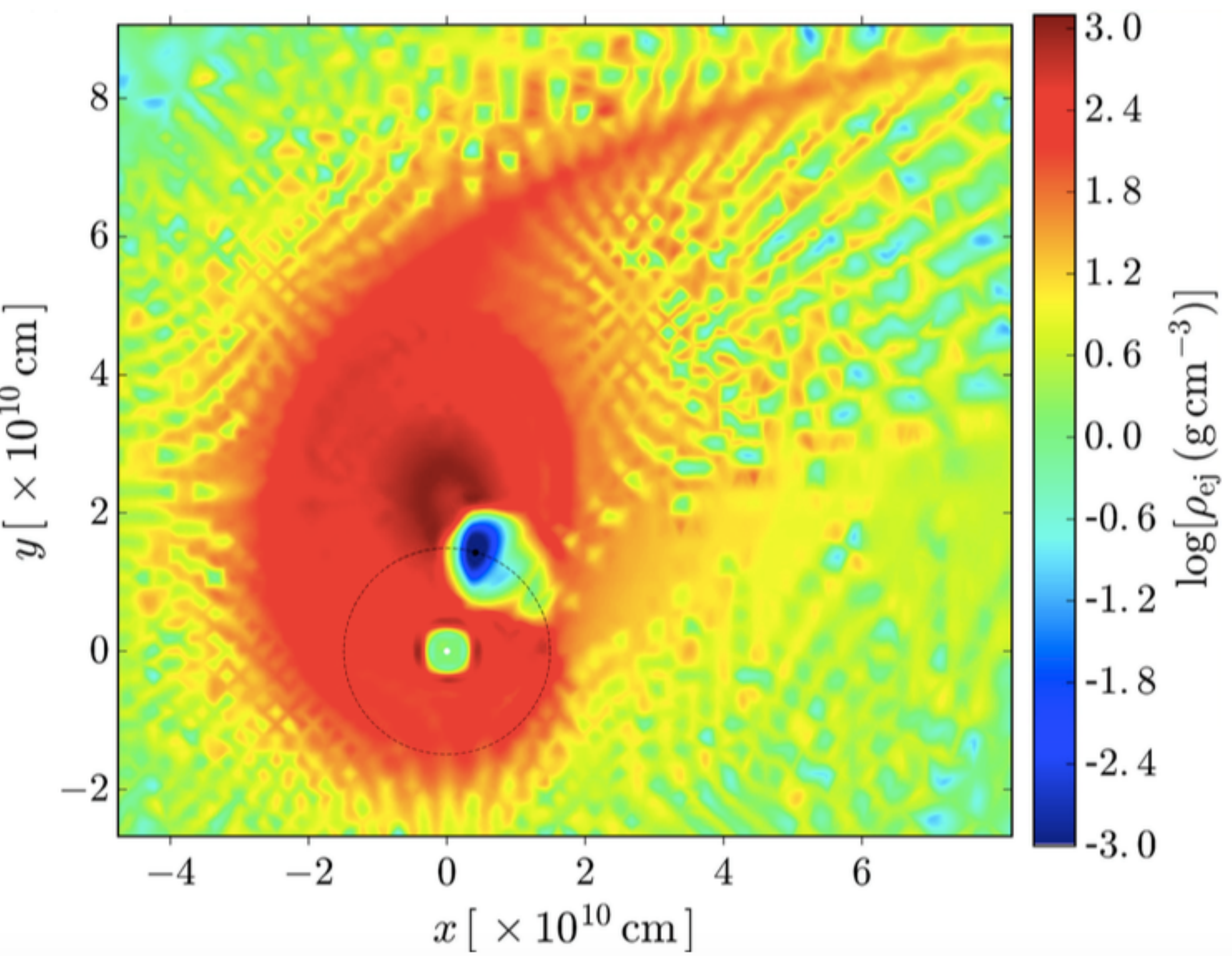}
\caption{The distribution of matter in the equatorial plane of the binary progenitors corresponding to the BH formation leading to a large cavity of radius of $\approx 10^{11}$~cm created in the HN ejecta. The blue color represents this cavity of the low density of $\approx 10^{-3}$~g~cm$^{-3}$ around the BH. The plot is taken from \citep{2016ApJ...833..107B} with the author's permission.  }\label{beccera}
\end{figure*}

Episode 1 corresponds to a ``\textit{precursor}'' which includes 1) the  thermal emission of 0.32 s originating from the SN shockwave breakout of the CO$_{\rm core}$ collapse (see companion paper Li et al. 2019, submitted for publication) 2) the subsequent hypercritical accretion of the SN ejecta onto the companion NS and 3) the reaching of the critical mass of the NS leading to the formation of a BH. This episode takes place between the \textit{Fermi}-GBM trigger rest frame time $t_{\rm rf}=0$~s and $~t_{\rm rf}=1.9$~s.  It encompasses 36$\%$ of the  total energy observed by the GBM, corresponding to an isotropic energy of $E_{\rm iso}=(1.0 \pm 0.12) \times 10^{53}$ erg. It contains the thermal component from $t_{\rm rf}=0.79$ s to $t_{\rm rf}=1.18$ s, with an isotropic energy of $E_{\rm sh}=(2.82 \pm 0.13) \times 10^{52}$~erg. We compare and contrast the duration, the shockwave break energy $E_{sh}$, and the isotropic energy $E_{iso}$ with three additional BdHNe I: GRB 130427A, 160509A, and 160625B.  All the results are reported in Table~\ref{thermal} and details are given in the companion article (Li et al., 2019, submitted for publication). Particular attention is given to make the comparison in the rest frame of the source, and list the corresponding redshift.  Interesting correlations exist between the energy of the shockwave breakout and the co-rotational kinetic energy of the CO core in the binary system, strictly related to the binary separation, binary period and total mass of the system, details are presented in Li et al. (2019, Submitted for publication).

The self enclosure inside the BH horizon of the companion NS and of the accreted material depletes the BdHN, as originally described in \cite{2016ApJ...833..107B,2019ApJ...871...14B}, by approximately $10^{57}$ baryons creating  a large cavity of radius $\approx 10^{11}$~cm in the hypernova (HN) ejecta around the BH site, see Fig.~\ref{beccera}. The occurrence of this cavity creates the condition of low baryon density necessary to observe the higher energy emission which occurs in the following two episodes.

\begin{figure*}
\centering
\includegraphics[angle=0, scale=0.6]{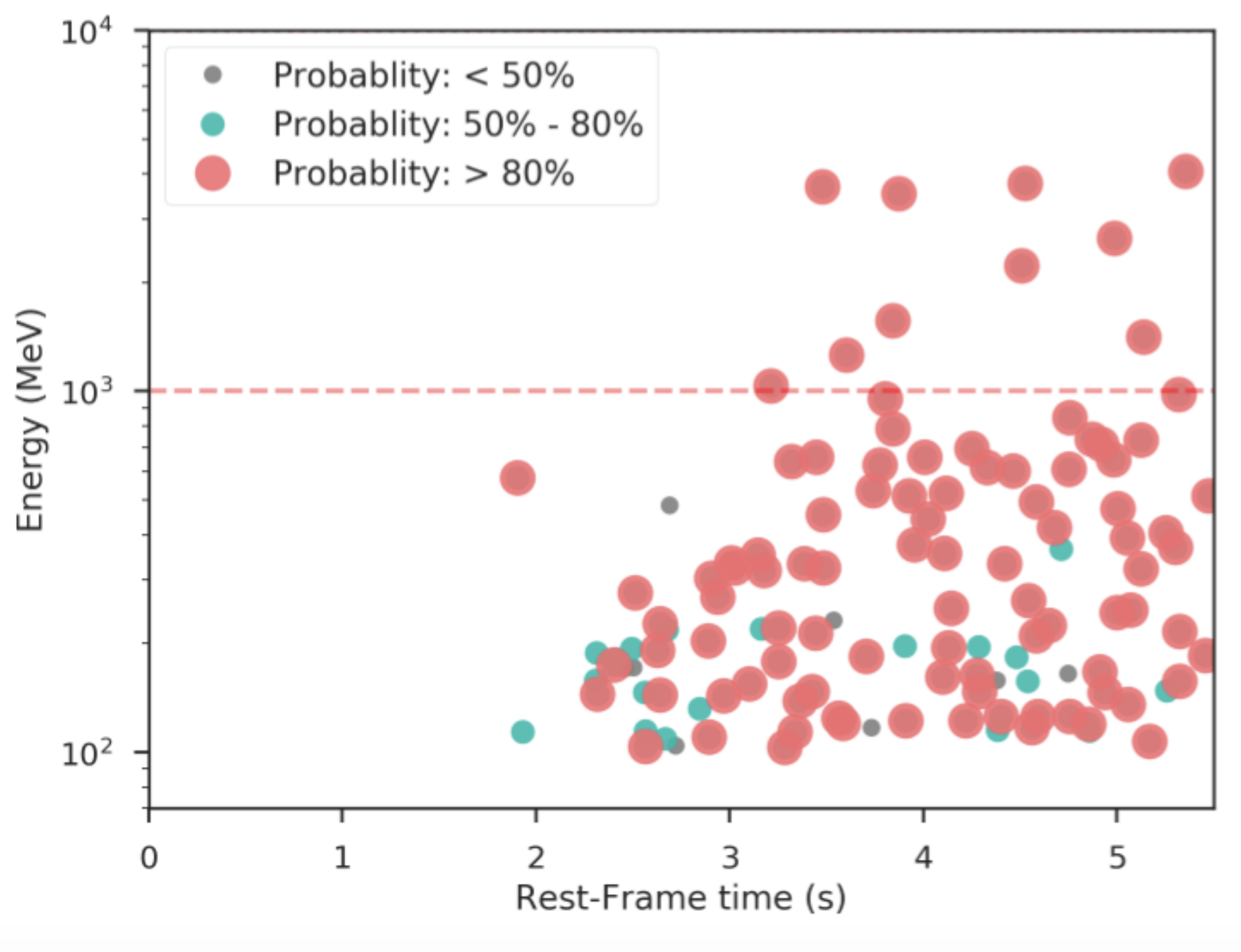}
\caption{In the first $t_{\rm rf}=$1.9~s there is no confident statistical evidence for  the GeV emission. The first confident GeV photon arrives at $t_{\rm rf}=$1.9~s with high probability of more than 80$\%$ belonging to the GRB 190114C. For more information see \citep{2019arXiv190107505W}.}\label{firstGeV}
\end{figure*}

Episode 2, the \textit{main spike} of the GRB energy emission lasting only 2 seconds, encompasses 59$\%$  of the energy observed by Fermi-GBM with an equivalent isotropic energy of $E_{\rm iso}=(1.47 \pm 0.2) \times 10^{53}$~erg. A similar spike has already been observed in GRB 151027A, \citep[see Fig.~4 in][]{2018ApJ...869..151R}. It occurs from $t_{\rm rf}=1.9$~s to $~t_{\rm rf}=3.99$~s. This episode is itself characterized by the simultaneous occurrence at $t_{\rm rf}=1.9 ~$s of three major events: 

a) the formation of the BH; details in section~\ref{sec:innerengine}.

b) The onset of the GeV emission signed by the observation of Fermi--LAT of the first GeV photon in the range $0.1$--$100$~GeV, see Fig.~\ref{firstGeV}. The total energy emitted by this source in the above GeV range is $E_{\rm GeV}=(1.8 \pm 0.9) \times 10^{53}$~erg \citep{2019arXiv190107505W}, which is comparable to the energy observed by the GBM. The GeV luminosity follows the power-law
\begin{equation}\label{L1}
    L = A~t^{-\alpha} \textrm{erg/s},
\end{equation}
with amplitude $A = (7.75\pm 0.44) \times 10^{52}$ and a slope of index of $\alpha = 1.2 \pm 0.36$, in agreement with the one observed in GRB 130427A \citep{2018arXiv181200354R}, see also in the companion paper  (Ruffini, Moradi et al. 2019, in preparation).

c) The onset of the UPE in gamma-rays observed by Fermi-GBM. Its spectral analysis clearly identify a cutoff power-law (CPL) and a black body (BB) component as the best fit of the spectrum (CPL+BB), see Fig.~\ref{onebin}. The nature of this spectrum calls attention to the vacuum polarization process occurring in a Kerr-Newmann BH with overcritical electric field originating $10^{21}$~eV particles \citep{1975PhRvL..35..463D} to the self acceleration process of $e^+ e^-$ in an optically thick plasma engulfed with baryons, the PEMB pulse \citep{1999A&AS..138..511R,1999A&A...350..334R, 2000A&A...359..855R}, as well as to the concept of dyadosphere and dyadotorus \citep[see e.g.][and references therein]{2010PhR...487....1R}.

Episode 3 starts at $~t_{\rm rf}=11$~s and ends at $~t_{\rm rf}=20$~s. It encompasses the remaining 5$\%$ of the  entire GBM energy, see Fig.~\ref{thermal}. No black body component is present in its spectrum, which appears to be featureless, and is very similar to the second spike in GRB 151027A \citep{2018ApJ...869..151R}. 

Performing relativistic hydrodynamic (RHD) simulations of the interaction of the cavity with the $e^+e^-$ plasma we have shown in the companion paper \citep{RuffiniFuksman2019} that the GBM detection in this episode originates within the cavity generated around the BH by its formation process, see Fig.~\ref{beccera}. The cavity is then further depleted in its baryonic content by the action of the UPE and the GeV emission, creating the necessary condition for generating e.g. the TeV emission announced by MAGIC radiation \citep{GCN23701}.

From all the above we conclude that the determination of the redshift $z$ and the observation of the UPE by the GBM data, lasting only $2$~s in the present case of GRB 190114C, are sufficient to identify the long GRB 190114C as a BdHN I source, the later observations of the GeV emission and the afterglow been just necessary implied by the BdHN I model (see e.g. in the companion paper Ruffini, Moradi et al., 2019, in preparation).

\section{Self-Similarity and power laws in the UPE}\label{sec:self}

We now turn to the new result obtained by a time resolved analysis  of the UPE spectrum. 

\begin{figure*}
\centering
\includegraphics[angle=0, scale=0.80]{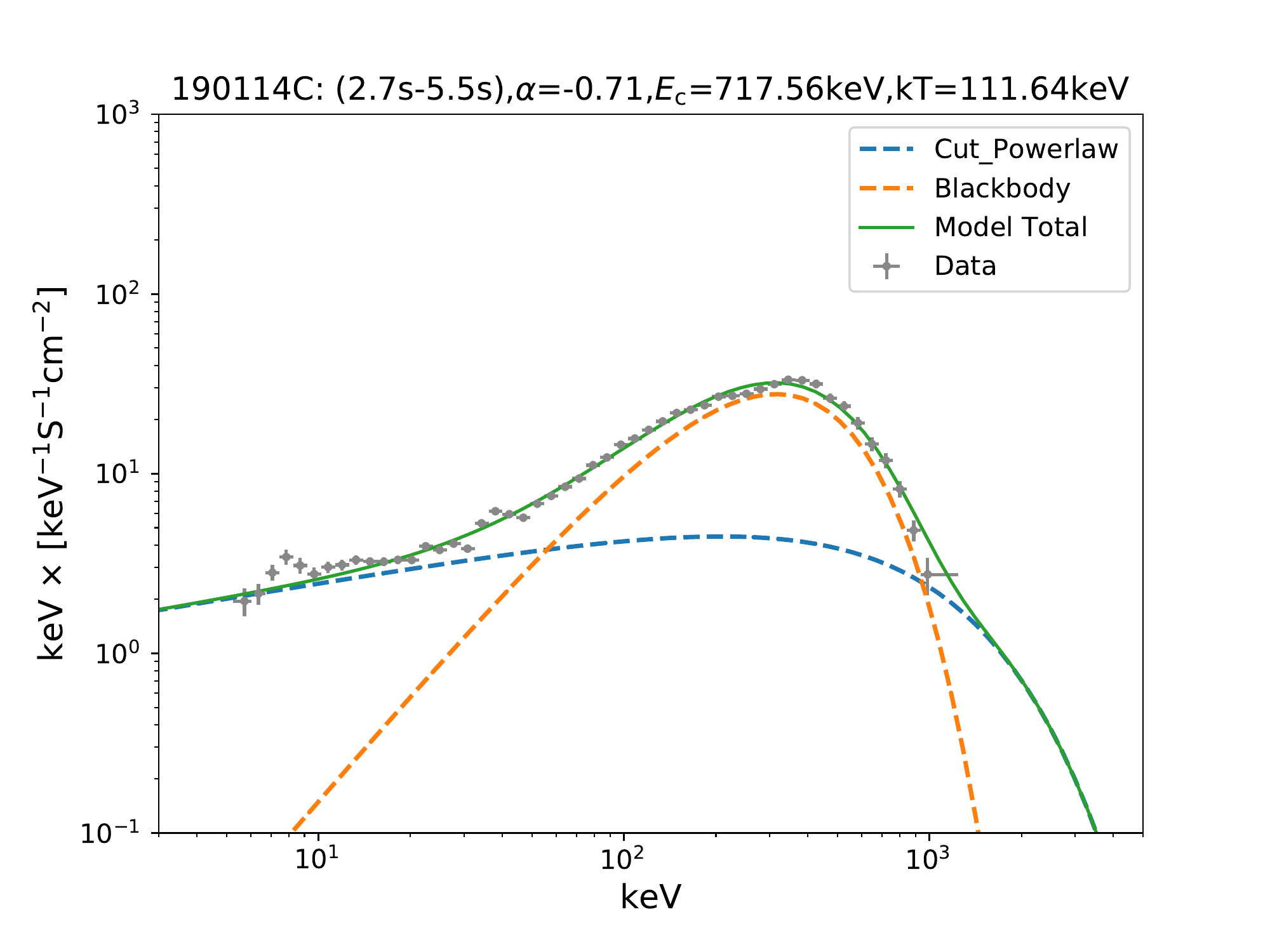}
\caption{The spectrum analysis for the time interval of $t_{\rm rf}=1.9$~s to $t_{\rm rf}=3.9$~s. We find the CPL+BB model is the preferred model as compared to all the other models. The best fit parameters of the power-law index of the CPL component, $\alpha= -0.71^{+0.02}_{-0.02}$, the cut-off energy of the CPL component, $E_c=717.6^{+25.4}_{-25.4}$, the temperature of the  black body component, $kT= 159.0^{+3.6}_{-3.6}$ keV, the likelihood -log(posterior)/AIC/BIC with the values of -3344/6697/6719, the  flux of black body component, $F_{\rm BB}=22.49^{+3.21}_{-2.65}$ ( $10^{-6}$~erg~cm$^{-2}$~s$^{-1}$) and the total flux of CPL+black body, $F_{\rm Total}=$ 111.10$^{+11.60}_{-10.40}$ ($10^{-6}~$erg~cm$^{-2}$~s$^{-1}$) are reported in Table.~\ref{tab:table}. The ratio of black body flux to the total flux, $F_{\rm BB}/F_{\rm Total}=0.2$ and  the isotropic energy of this time interval $E_{\rm iso}=1.5 \times 10^{53}$~erg. 
}\label{onebin} 
\end{figure*}

\begin{figure*}
\centering
\includegraphics[angle=90, width=0.84\hsize]{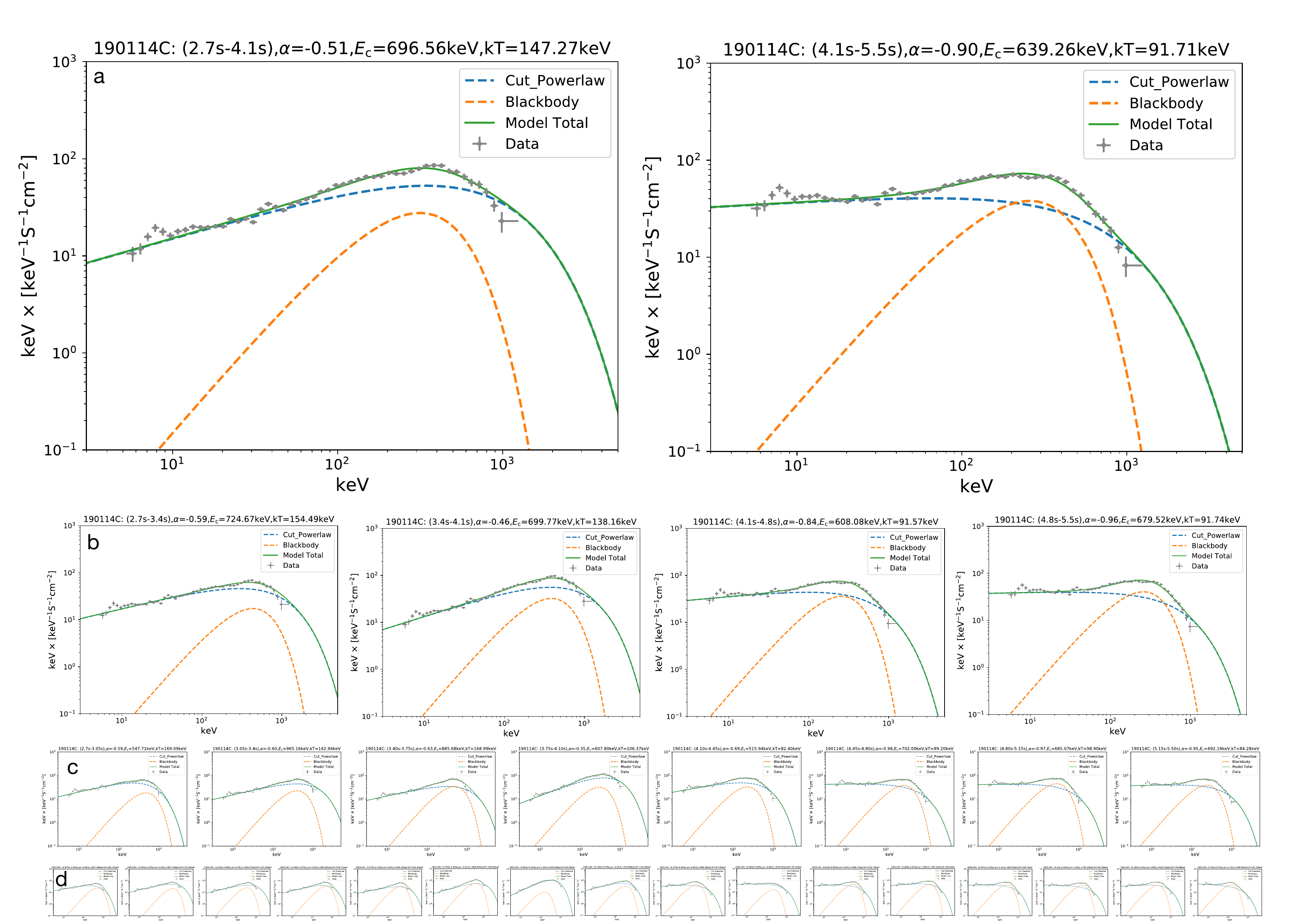}
\caption{The same as Fig.~\ref{onebin}, but the time interval is divided into two equal parts (top line), four equal parts (second line), eight equal parts (third line), and sixteen equal parts (bottom line), respectively.}\label{alltogether}
\end{figure*}

Following the spectral analysis performed over the entire time interval of $t_{\rm rf}= 1.9$~s to $t_{\rm rf}=3.9$~s reported in Fig.~\ref{onebin}, we divide the rest frame time interval in half and perform again the same spectral analysis for the two one second intervals, namely [$1.9$s--$2.9$s] and [$2.9$s--$3.9$s], obtaining the results shown in Fig.~\ref{alltogether}a.

We then divide each of these half intervals again in half, i.e., $\Delta t_{\rm rf}=0.5$~s corresponding to [$1.9$s--$2.40$s], [$2.40$s--$2.9$s], [$2.9$s--$3.4$s] and [$3.4$s--$3.9$s] and redo the previous spectral analysis obtaining the results in Fig.~\ref{alltogether}b.

In a fourth iteration we divide the UPE into $8$ sub-intervals of $\Delta t_{\rm rf}= 0.25$~s corresponding to the time intervals [$1.9$s--$2.15$s],[$2.15$s--$2.40$s], [$2.40$s--$2.65$s], [$2.65$s--$2.9$s], [$2.9$s--$3.15$s], [$3.15$s--$3.4$s], [$3.4$s--$3.65$s] and [$3.65$s--$3.9$s], 
and redo the spectral analysis, see Fig.~\ref{alltogether}c.

In the fifth and final iteration  of this process we divide the UPE into $16$ sub-intervals of $\Delta t_{\rm rf}= 0.125$~s corresponding we perform the spectral analysis and find the self-similar CPL+BB emission in  the time intervals [$2.019$s--$2.142$s],[$2.142$s--$2.265$s], [$2.265$s--$2.388$s], [$2.388$s--$2.511$s], [$2.511$s--$2.633$s], [$2.633$s--$2.756$s], [$2.756$s--$2.87$s], [$2.879$s--$3.002$s], [$3.002$s--$3.125$s], [$3.125$s--$3.248$s], [$3.248$s--$3.371$s], [$3.371$s--$3.494$s], [$3.494$s--$3.617$s], [$3.617$s--$3.739$s], [$3.739$s--$3.862$s] and [$3.862$s--$3.985$s] and perform the spectral analysis, see Fig.~\ref{alltogether}d.

The results of these spectral fittings are shown in Fig.~\ref{onebin}, Fig.~\ref{alltogether} and Table~\ref{tab:table}.
The reported self-similarity in Table~\ref{tab:table} has led to three main new results:

\textit{The first} has allowed the identification with much greater precision of the onset and end time of the presence of the CPL+BB spectra in Fig.~\ref{alltogether} in the fifth iterative step. Consequently, the onset time of the self-similar structure is $t_{\rm rf}=2.01$~s and the end time is $t_{\rm rf}=3.99$~s.

\begin{figure*}
\centering
\includegraphics[angle=0, scale=0.53]{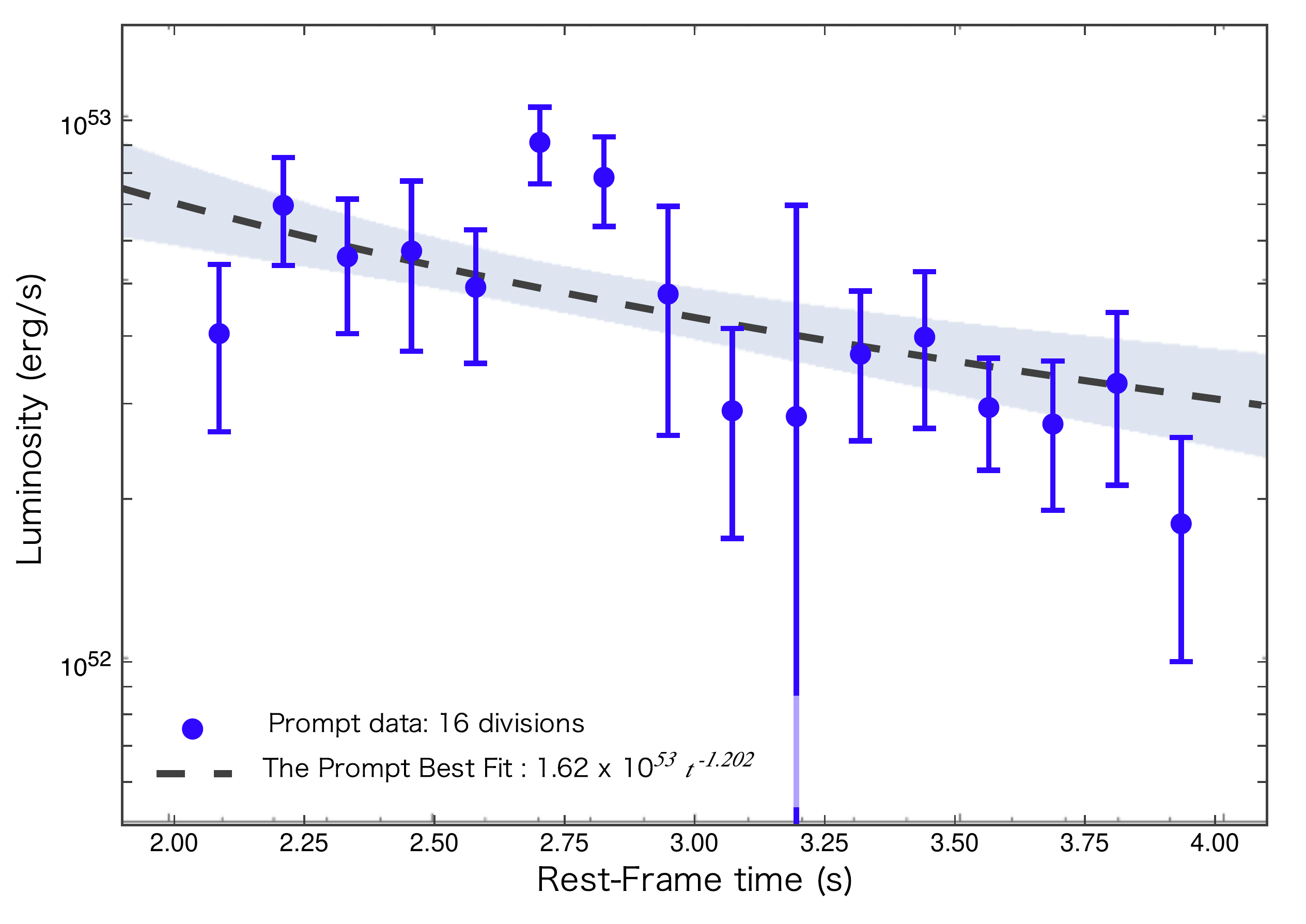}
\includegraphics[angle=0, scale=0.53]{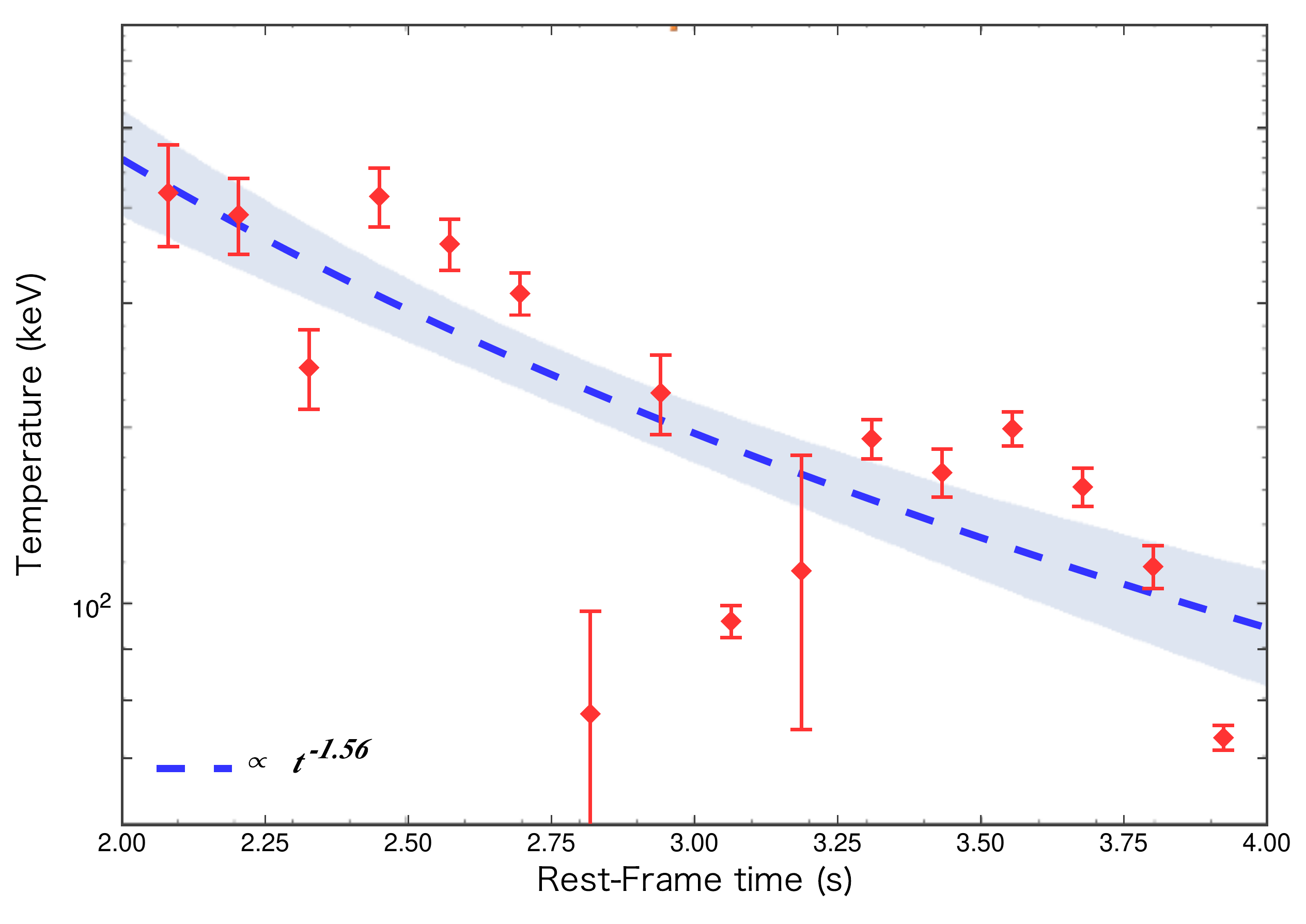}
\caption{\textbf{Upper}: Time evolution of the luminosity of the UPE as derived from the fifth iteration with 16 sub-intervals. The values of the  best fit parameters from Table.~\ref{tab:table} are used to apply the k-correction and measuring the luminosity as a function of time. The power law index of $-1.20 \pm 0.26$ of the luminosity is similar to the one obtained in the Gev emission luminosity with index of $-1.20 \pm 0.36$, see Eq.~\ref{L1}, Eq.~\ref{L2} and Eq.~\ref{L3}. For more information about GeV luminosity behavior see \citep{2019arXiv190107505W} and the companion paper (Ruffini, Moradi et al, 2019, in preparation). \textbf{Lower}: Time evolution of the rest-frame temperature of the UPE as derived from the fifth iteration with $16$ sub-intervals as reported in Table~\ref{tab:table}.
}\label{iterativ}
\end{figure*}

\textit{The second} is for each sample in the iterative step we have identified the value of the power-law cutoff, the cutoff energy, the total energy, the corresponding black body temperature, the black body and the total flux and the total energy, see Table~\ref{tab:table} and Fig.~\ref{iterativ}. The ratio of black body flux to the total flux, $F_{\rm BB}/F_{\rm Total}$, remains constant in each sample, see column of $F_{ratio}$ in table ~\ref{tab:table}.

\textit{The third} is that the total flux and parameters of the best fit, when the k-correction is applied, allows the estimation of the luminosity as a function of time. The inferred luminosity derived from the crude approximation of the fifth iteration into $16$ sub-intervals indicates a power-law luminosity, measured in the rest frame,
\begin{equation}\label{L2}
    L = A~ t^{-\alpha} \textrm{erg/s},
\end{equation}
with amplitude $A = (1.62\pm 0.77) \times 10^{53}$ and a slope of index of $\alpha = 1.2 \pm 0.26$ consistent with the one measured in the GeV radiation, see Fig.~\ref{iterativ} and results in \citet{2019arXiv190107505W}, \citet{2018arXiv181200354R} and Ruffini, Moradi et al, 2019 (in preparation). Similarly, a power-law with an index of $-1.56$ is found for the temperature, see Fig.~\ref{iterativ}. 

In the above sequence it appears that the self-similarity continues rigorously in each successive time iteration, with the only possible limitations being established by the sensitivity of the observations and the capability of the satellite to perform the observations with the necessary precision. Our examination shows that no evidence for any intrinsic departure from self-similarity can be found from the available observational data.

The natural question then arises: what is the deeper origin of this self-similarity and at what level will the self-similarity stop revealing the basic nature of the ``\textit{inner engine}'' which manifests itself in the occurrence of this self-similar structure?

\begin{deluxetable*}{cccccccccc}
\tabletypesize{\tiny}
\tablecaption{Results of the time-resolved Spectral fits of GRB 190114C (CPL+BB model) from $t_{\rm rf}=1.9$~s to $t_{\rm rf}=3.99$~s. The time intervals both in the rest-frame and observer frame, the power-law index, cut-off energy, temperature, AIC/BIC, BB flux, total flux, the ratio of black body flux to the total flux, $F_{\rm BB}/F_{\rm Total}$ and finally the isotropic energy are reported in this table. The $F_{\rm BB}/F_{\rm Total}$ remains almost constant in each sample. The Akaike Information Criterion (AIC, \citealt{1974ITAC...19..716A}) and the Bayesian Information Criterion (BIC, \citealt{schwarz1978estimating}) can be used to select non-nested models. The AIC and BIC are defined as AIC=-2ln$L(\vec{\theta})$+2k and BIC=-2ln$L(\vec{\theta})$+kln(n), respectively. Here $L$ is the maximized value of the likelihood function for the estimated model, $k$ is the number of free parameters to be estimated, $n$ is the number of observations (or the sample size). The prefer model between any two estimated models is the one that provides the minimum AIC and BIC scores. After comparing the AIC and BIC , we find the CPL+BB model is the preferred model than the CPL  and other model. The likelihood -log(posterior) and the AIC and BIC scores are reported in column 6. There is in the fifth iteration  a delay of $0.1$~s between the onset of the GeV radiation and the onset of the UPE.\label{tab:table}}
\tablehead{ 
\colhead{$t_{1}$$\sim$$t_{2}$}
&\colhead{$t_{1}$$\sim$$t_{2}$}
&\colhead{$\alpha$}
&\colhead{$E_{\rm c}$}
&\colhead{kT}
&\colhead{-log(posterior)/(AIC/BIC)}
&\colhead{$F_{\rm BB}$}
&\colhead{$F_{\rm Total}$}
&\colhead{$F_{\rm ratio}$} 
&\colhead{$E_{\rm Total}$}\\
\hline
(s)&(s)&&(keV)&(keV)&&(10$^{-6}$)&(10$^{-6}$)&&(erg)\\
Observation&Rest&&&Rest&&(erg cm$^{-2}$ s$^{-1}$)&(erg cm$^{-2}$ s$^{-1}$)&&Rest}
\colnumbers
\startdata
2.700$\sim$5.500&1.896$\sim$3.862&-0.71$^{+0.02}_{-0.02}$&717.6$^{+25.4}_{-25.4}$&159.0$^{+3.6}_{-3.6}$&-3344/6697/6719&22.49$^{+3.21}_{-2.65}$&111.10$^{+11.60}_{-10.40}$&0.20&1.50e+53\\
\hline
2.700$\sim$4.100&1.896$\sim$2.879&-0.51$^{+0.02}_{-0.02}$&696.6$^{+31.9}_{-32.4}$&209.7$^{+9.3}_{-9.1}$&-2675/5360/5381&24.67$^{+6.93}_{-5.35}$&142.50$^{+23.90}_{-21.00}$&0.17&9.64e+52\\
4.100$\sim$5.500&2.879$\sim$3.862&-0.90$^{+0.02}_{-0.02}$&639.3$^{+31.9}_{-31.6}$&130.6$^{+2.5}_{-2.5}$&-2529/5069/5090&25.55$^{+2.97}_{-2.75}$&80.98$^{+9.68}_{-8.07}$&0.32&5.48e+52\\
\hline
2.700$\sim$3.400&1.896$\sim$2.388&-0.59$^{+0.03}_{-0.03}$&724.7$^{+44.5}_{-45.5}$&220.0$^{+17.1}_{-17.2}$&-1882/3774/3796&18.55$^{+9.42}_{-7.40}$&123.90$^{+29.20}_{-22.30}$&0.15&4.19e+52\\
3.400$\sim$4.100&2.388$\sim$2.879&-0.46$^{+0.04}_{-0.04}$&699.8$^{+47.8}_{-48.3}$&196.7$^{+8.9}_{-8.7}$&-2032/4074/4095&31.78$^{+9.60}_{-7.31}$&161.40$^{+47.10}_{-32.40}$&0.20&5.46e+52\\
4.100$\sim$4.800&2.879$\sim$3.371&-0.84$^{+0.03}_{-0.03}$&608.1$^{+42.1}_{-42.2}$&130.4$^{+3.7}_{-3.9}$&-1880/3770/3792&23.94$^{+4.20}_{-4.22}$&85.37$^{+14.83}_{-12.27}$&0.28&2.89e+52\\
4.800$\sim$5.500&3.371$\sim$3.862&-0.96$^{+0.03}_{-0.03}$&679.5$^{+49.1}_{-48.7}$&130.6$^{+3.1}_{-3.2}$&-1809/3628/3649&27.18$^{+4.01}_{-3.73}$&78.20$^{+11.40}_{-9.66}$&0.35&2.65e+52\\
\hline
2.700$\sim$3.050&1.896$\sim$2.142&-0.59$^{+0.03}_{-0.03}$&547.7$^{+44.2}_{-44.9}$&240.8$^{+29.2}_{-29.1}$&-1187/2384/2406&19.67$^{+17.96}_{-8.88}$&103.20$^{+30.60}_{-20.28}$&0.19&1.75e+52\\
3.050$\sim$3.400&2.142$\sim$2.388&-0.60$^{+0.02}_{-0.02}$&965.2$^{+28.5}_{-30.1}$&203.5$^{+14.8}_{-14.8}$&-1320/2650/2671&22.87$^{+8.88}_{-7.23}$&152.00$^{+24.00}_{-21.00}$&0.15&2.57e+52\\
3.400$\sim$3.750&2.388$\sim$2.633&-0.63$^{+0.04}_{-0.04}$&885.7$^{+70.9}_{-70.1}$&240.6$^{+10.5}_{-10.6}$&-1224/2458/2480&41.02$^{+11.09}_{-7.91}$&129.10$^{+32.40}_{-23.40}$&0.32&2.18e+52\\
3.750$\sim$4.100&2.633$\sim$2.879&-0.35$^{+0.06}_{-0.05}$&607.8$^{+57.1}_{-60.1}$&151.5$^{+12.4}_{-14.2}$&-1428/2866/2887&23.92$^{+12.46}_{-10.40}$&192.00$^{+101.70}_{-60.30}$&0.12&3.25e+52\\
4.100$\sim$4.450&2.879$\sim$3.125&-0.69$^{+0.04}_{-0.04}$&515.9$^{+43.6}_{-43.6}$&117.3$^{+5.0}_{-5.0}$&-1271/2552/2573&19.19$^{+4.89}_{-4.40}$&92.71$^{+27.69}_{-22.43}$&0.21&1.57e+52\\
4.450$\sim$4.800&3.125$\sim$3.371&-0.98$^{+0.04}_{-0.04}$&702.0$^{+78.1}_{-78.2}$&141.3$^{+5.8}_{-5.8}$&-1254/2518/2539&26.76$^{+6.41}_{-5.47}$&80.73$^{+17.95}_{-14.95}$&0.33&1.37e+52\\
4.800$\sim$5.150&3.371$\sim$3.617&-0.97$^{+0.04}_{-0.04}$&685.1$^{+69.4}_{-68.6}$&140.8$^{+4.6}_{-4.6}$&-1218/2447/2468&31.83$^{+6.85}_{-4.98}$&82.51$^{+15.62}_{-12.33}$&0.39&1.40e+52\\
5.150$\sim$5.500&3.617$\sim$3.862&-0.95$^{+0.04}_{-0.04}$&692.2$^{+79.1}_{-77.7}$&120.0$^{+4.0}_{-4.0}$&-1203/2416/2438&23.19$^{+5.38}_{-3.81}$&73.57$^{+18.69}_{-12.93}$&0.32&1.24e+52\\
\hline
2.875$\sim$3.050&2.019$\sim$2.142&-0.68$^{+0.04}_{-0.05}$&627.6$^{+87.0}_{-91.5}$&258.0$^{+30.1}_{-28.7}$&-664/1337/1359&28.45$^{+20.42}_{-12.51}$&98.14$^{+33.56}_{-26.44}$&0.29&8.30e+51\\
3.050$\sim$3.225&2.142$\sim$2.265&-0.59$^{+0.03}_{-0.03}$&957.1$^{+34.1}_{-34.9}$&245.3$^{+21.5}_{-21.0}$&-768/1547/1568&25.71$^{+13.87}_{-9.03}$&169.30$^{+38.20}_{-31.60}$&0.15&1.43e+52\\
3.225$\sim$3.400&2.265$\sim$2.388&-0.59$^{+0.03}_{-0.03}$&929.6$^{+55.7}_{-58.7}$&172.4$^{+15.8}_{-15.8}$&-759/1527/1549&20.60$^{+11.17}_{-8.22}$&136.00$^{+37.90}_{-26.70}$&0.15&1.15e+52\\
3.400$\sim$3.575&2.388$\sim$2.511&-0.59$^{+0.05}_{-0.05}$&804.0$^{+86.7}_{-82.3}$&255.9$^{+17.4}_{-17.4}$&-702/1414/1436&42.19$^{+19.41}_{-13.59}$&139.30$^{+48.30}_{-35.60}$&0.30&1.18e+52\\
3.575$\sim$3.750&2.511$\sim$2.633&-0.65$^{+0.04}_{-0.04}$&916.3$^{+64.6}_{-67.7}$&229.3$^{+13.6}_{-13.5}$&-730/1471/1492&39.25$^{+11.97}_{-10.71}$&119.50$^{+32.90}_{-25.45}$&0.33&1.01e+52\\
3.750$\sim$3.925&2.633$\sim$2.756&-0.51$^{+0.02}_{-0.02}$&960.9$^{+30.9}_{-31.4}$&204.6$^{+9.9}_{-10.0}$&-808/1627/1648&57.70$^{+15.81}_{-12.25}$&221.10$^{+35.60}_{-31.50}$&0.26&1.87e+52\\
3.925$\sim$4.100&2.756$\sim$2.879&-0.19$^{+0.04}_{-0.04}$&416.6$^{+26.0}_{-26.3}$&77.5$^{+20.7}_{-16.1}$&-818/1646/1668&4.29$^{+6.85}_{-3.22}$&177.30$^{+57.10}_{-45.70}$&0.02&1.50e+52\\
4.100$\sim$4.275&2.879$\sim$3.002&-0.54$^{+0.06}_{-0.06}$&474.1$^{+45.5}_{-46.2}$&162.6$^{+14.9}_{-14.8}$&-758/1526/1547&24.26$^{+17.09}_{-10.09}$&116.10$^{+52.40}_{-35.12}$&0.21&9.82e+51\\
4.275$\sim$4.450&3.002$\sim$3.125&-0.90$^{+0.06}_{-0.06}$&586.4$^{+86.7}_{-89.0}$&96.0$^{+3.5}_{-3.5}$&-738/1485/1507&21.95$^{+4.75}_{-4.49}$&70.69$^{+29.51}_{-17.13}$&0.31&5.98e+51\\
4.450$\sim$4.625&3.125$\sim$3.248&-0.96$^{+0.20}_{-0.10}$&679.6$^{+148.4}_{-182.5}$&107.9$^{+33.0}_{-94.9}$&-722/1454/1475&16.62$^{+12.12}_{-16.62}$&68.87$^{+100.23}_{-28.45}$&0.24&5.82e+51\\
4.625$\sim$4.800&3.248$\sim$3.371&-0.95$^{+0.05}_{-0.05}$&694.2$^{+96.8}_{-94.2}$&146.3$^{+6.7}_{-6.6}$&-734/1477/1499&35.59$^{+9.47}_{-8.00}$&89.91$^{+27.59}_{-18.82}$&0.40&7.60e+51\\
4.800$\sim$4.975&3.371$\sim$3.494&-0.85$^{+0.05}_{-0.05}$&564.5$^{+68.9}_{-71.9}$&135.3$^{+7.5}_{-7.6}$&-744/1498/1519&30.78$^{+11.12}_{-8.55}$&96.58$^{+31.02}_{-23.68}$&0.32&8.17e+51\\
4.975$\sim$5.150&3.494$\sim$3.617&-1.10$^{+0.04}_{-0.04}$&820.5$^{+115.0}_{-111.2}$&149.7$^{+5.9}_{-5.8}$&-683/1376/1398&32.76$^{+6.98}_{-5.92}$&71.57$^{+16.74}_{-11.99}$&0.46&6.05e+51\\
5.150$\sim$5.325&3.617$\sim$3.739&-1.04$^{+0.05}_{-0.05}$&765.2$^{+119.0}_{-115.8}$&130.9$^{+5.8}_{-5.8}$&-697/1404/1426&26.14$^{+7.02}_{-5.96}$&66.70$^{+20.48}_{-14.17}$&0.39&5.64e+51\\
5.325$\sim$5.500&3.739$\sim$3.862&-0.88$^{+0.06}_{-0.06}$&635.3$^{+88.7}_{-92.0}$&108.9$^{+5.3}_{-5.4}$&-736/1483/1504&20.90$^{+6.51}_{-5.15}$&79.48$^{+28.02}_{-21.03}$&0.26&6.72e+51\\
5.500$\sim$5.675&3.862$\sim$3.985&-1.10$^{+0.08}_{-0.08}$&568.5$^{+130.8}_{-125.4}$&73.4$^{+2.1}_{-2.1}$&-657/1324/1345&16.08$^{+2.77}_{-2.56}$&43.59$^{+19.37}_{-10.92}$&0.37&3.69e+51\\
\enddata
\end{deluxetable*}

%

\section{\emph{Inner engine} properties}\label{sec:innerengine}

In recent months we have addressed the underlying physical process originating  the GeV emission and its observed power-law luminosity using as prototype GRB 130427A \citep{2018arXiv181101839R, 2018arXiv181200354R}, the twin source of GRB 190114C. We have addressed the outstanding problem of relativistic astrophysics of how to extract the rotational energy from a Kerr BH for powering the synchrotron emission in the observed GeV luminosity. We have there identified as the ``inner engine'', for GRB 130427A, a Kerr BH  of initial mass $M=2.285 M_{\odot}$, and angular momentum per-unit-mass $a = J/M = 0.303 M$, where $J$ is the BH angular momentum, in presence of a background uniform magnetic field of $10^{14}$~G aligned with the BH rotation axis, a well known solution of the Einstein equations mathematically derived by \citet{Wald:1974np}. The initial impulsive emission process occurs assuming that at the BH horizon the induced electric field of the Wald solution is critical, i.e. $E = E_c$ where
\begin{equation}\label{eq:Ec}
 E_{c}=\frac{m_e^2 c^3}{e\hbar}\,
\end{equation}
with $m_e$ and $e$ the electron mass and charge, respectively.

The corresponding energy extraction process occurs in an impulsive process in the regime of under-critical electric field, $E < E_{c}$. In the first process the electric potential difference $\Delta\phi$ can accelerate each proton up to $\epsilon_{p} = e \Delta\phi  \approx 10^{21}$~eV. The total energy of $\mathcal{E}\approx 10^{44}$~erg accelerates $N_{p}= \mathcal{E}/\epsilon_{p} \approx 10^{34}$ protons in a theoretically derived time scale of 
\begin{equation}
 \tau_{\rm th}=\Delta\phi/(c E) = r_+/c \approx 10^{-6}~\textrm{s},
\end{equation}
where $r_+$ is the BH horizon radius, leading to a GeV luminosity consistent with the observation, see \citep{2018arXiv181101839R}.

In a further extension of this work \citep{2018arXiv181200354R}, we have  derived how the ultra-relativistic protons at $10^{21}$~eV when propagating along the polar axis, for an injection angle $\theta=0$, give origin to ultra high-energy cosmic rays (UHECRs). When propagating with $\theta \neq 0$, they give origin to synchrotron emission in the GeV, TeV and PeV energies in a theoretically computed acceleration time independent of the emitted photon energy 
\begin{equation}
 \tau_{\rm th}= 3 \times 10^{-6}~\textrm{s},   
\end{equation}
which is, however, a function of circumburst medium density, details in \citep{2018arXiv181101839R}.

In view of the absence of the detailed GBM data in GRB 130427A recalled above, we have there conservatively started the evaluation of the repetitive process at $37$~s.
 
Correspondingly, we have determined the sequence of the impulsive  process which with an observed characteristic repetition time of
 \begin{equation}
 \tau_{\rm obs}= \frac{\mathcal{E}}{L}= 1.2 \times 10^{-6}~\textrm{s},   
\end{equation}
slowly increasing with the time evolution. In each impulsive event we have derived the decrease of the mass and spin of the BH necessary for powering the GeV emission. We conclude that this BdHN machine can in principal sustain the GeV emission for thousands of years \citep{2015ApJ...798...10R}.

The main conclusion is that in GRB 130427A the GeV emission observed \textit{macroscopically} to be emitted continuously with a luminosity
\begin{equation}\label{L3}
    L = A~ t^{-\alpha} \textrm{ erg/s}
\end{equation}
with amplitude $A = (2.05\pm 0.23) \times 10^{52}$ and a slope of index of $\alpha=1.2 \pm 0.04$,  when observed, in fact, \textit{microscopically} occurs in a sequence of elementary impulses each lasting $\approx 10^{-6}$ seconds.

Therefore, the emission of the GeV luminosity is microscopically a ``\textit{discrete process}'', compared and contrasted, macroscopically, to a ``\textit{continuous process}''. 

The GeV emission is composed of a series of $\approx 10^{6}$ ``\textit{discrete pulses}'' per second in GRB 130427A,  each with  an energy of $\approx 10^{44}$~erg and accelerating protons to $10^{21}$ eV \citep{2018arXiv181101839R}.

This treatment is being extended now to GRB 190114C.

\section{The generalization of GRB 130427A approach to GRB 190114C and the other BdHN I}\label{sec:generalization}

In the generalization of GRB 130427A approach to GRB 190114C and other BdHN I the basic assumption is adopted: that the \emph{inner engine}, during the UPE phase from $t_{\rm rf}=1.9$~s all the way to $t_{\rm rf}=3.99$~s, operates in the regime $E>E_{\rm c}$. For time following $3.99$~s the \emph{inner engine} operates in the regime $E<E_{\rm c}$.

Therefore $t_{\rm rf}=3.99$~s represents a separatrix between two different regimes in the UPE phase. From $1.9$~s to $3.99$~s two different process occur simultaneously:  the first based on classical electrodynamics already applied to GRB 130427A leading to the process of synchrotron radiation and to the generation of the GeV, TeV, PeV and UHECRs emission. The second based on quantum electrodynamics process of vacuum polarization needs a  different conceptual analysis based on the quantum phenomena tested in \citet{ 1999A&AS..138..511R,1999A&A...350..334R, 2000A&A...359..855R,2003PhLB..573...33R, 2010PhR...487....1R}. After $3.99$~s, only the classical electrodynamics remains active. 

We  address generally this new and challenging problematic in a companion paper (Ruffini, Moradi et al. 2019, in preparation): there we apply our micro-physical approach to describe the GeV emission in GRB 190114C, imposing $E \le E_c$ at $t_{\rm rf } \geq 3.99$~s, in agreement with the results presented in Table \ref{tab:table}.

This novel scenario is in front of us: we are currently verifying the presence of self-similarity and power-laws in the other two BdHNe I,  GRB 160625B and GRB 160502A and explore the extent of their validity and generality of our approach, see e.g., the companion paper Li et al., (2019)

\section{Conclusions}

From the data, of 1) the onset of the UPE, 2) the onset of the GeV radiation and 3) the onset of the \emph{inner engine}, there is no observational evidence which prevents considering these three events to be perfectly simultaneous. There are clear theoretical arguments for their temporal coincidence.

The observed self-similarities in the UPE phase as well as the sequence of \textit{discrete} impulsive emissions occurring in  time scales of $10^{-8}$--$10^{-6}$~s cannot be explained within the traditional model of a GRB based on a \textit{continuous} description of a relativistic blast wave, see e.g., \citet[][]{Meszaros:2001vr,2004RvMP...76.1143P}. 

We are opening a new avenue of research to exploit the concept of self-similarity revealed in the time-resolved spectra and in the power-law behavior in time, well established in the field of micro-physics in the works of Heisenberg \citep{1924AnP...379..577H}, Landau \citep{Landau:480039,Landau:480041} and Wilson \citep{RevModPhys.55.583},  here applied for the first time in the domain of GRBs, active galactic nuclei and relativistic astrophysics based on Einstein's theory of general relativity.  

\acknowledgments One of us (RR) acknowledges Carlo Di Castro for discussions.


\begin{thebibliography}{}
\expandafter\ifx\csname natexlab\endcsname\relax\def\natexlab#1{#1}\fi
\providecommand{\url}[1]{\href{#1}{#1}}

\bibitem[{{A. Melandri} {et~al.}(2019){A. Melandri}, D'Avanzo, D.~Malesani~and,
  Pian, N.~R. Tanvir~and, Olivares, Carini, Palazzi, Piranomonte, Jonker,
  Rossi, Kann, Hartmann, Inserra, Kankare, Maguire, Smartt, Yaron, Young,
  Manulis, \& on~behalf of a~larger collaboration}]{GCN23983}
{A. Melandri}, D'Avanzo, L. I.~P., D.~Malesani~and, M. D.~V., {et~al.} 2019,
  GRB Coordinates Network

\bibitem[{{Ackermann} {et~al.}(2014){Ackermann}, {Ajello}, {Asano}, {Atwood},
  {Axelsson}, {Baldini}, {Ballet}, {Barbiellini}, {Baring}, {Bastieri},
  {Bechtol}, {Bellazzini}, {Bissaldi}, {Bonamente}, {Bregeon}, {Brigida},
  {Bruel}, {Buehler}, {Burgess}, {Buson}, {Caliandro}, {Cameron}, {Caraveo},
  {Cecchi}, {Chaplin}, {Charles}, {Chekhtman}, {Cheung}, {Chiang}, {Chiaro},
  {Ciprini}, {Claus}, {Cleveland}, {Cohen-Tanugi}, {Collazzi}, {Cominsky},
  {Connaughton}, {Conrad}, {Cutini}, {D'Ammando}, {de Angelis}, {DeKlotz}, {de
  Palma}, {Dermer}, {Desiante}, {Diekmann}, {Di Venere}, {Drell},
  {Drlica-Wagner}, {Favuzzi}, {Fegan}, {Ferrara}, {Finke}, {Fitzpatrick},
  {Focke}, {Franckowiak}, {Fukazawa}, {Funk}, {Fusco}, {Gargano}, {Gehrels},
  {Germani}, {Gibby}, {Giglietto}, {Giles}, {Giordano}, {Giroletti}, {Godfrey},
  {Granot}, {Grenier}, {Grove}, {Gruber}, {Guiriec}, {Hadasch}, {Hanabata},
  {Harding}, {Hayashida}, {Hays}, {Horan}, {Hughes}, {Inoue}, {Jogler},
  {J{\'o}hannesson}, {Johnson}, {Kawano}, {Kn{\"o}dlseder}, {Kocevski}, {Kuss},
  {Lande}, {Larsson}, {Latronico}, {Longo}, {Loparco}, {Lovellette}, {Lubrano},
  {Mayer}, {Mazziotta}, {McEnery}, {Michelson}, {Mizuno}, {Moiseev}, {Monzani},
  {Moretti}, {Morselli}, {Moskalenko}, {Murgia}, {Nemmen}, {Nuss}, {Ohno},
  {Ohsugi}, {Okumura}, {Omodei}, {Orienti}, {Paneque}, {Pelassa}, {Perkins},
  {Pesce-Rollins}, {Petrosian}, {Piron}, {Pivato}, {Porter}, {Racusin},
  {Rain{\`o}}, {Rando}, {Razzano}, {Razzaque}, {Reimer}, {Reimer}, {Ritz},
  {Roth}, {Ryde}, {Sartori}, {Parkinson}, {Scargle}, {Schulz}, {Sgr{\`o}},
  {Siskind}, {Sonbas}, {Spandre}, {Spinelli}, {Tajima}, {Takahashi}, {Thayer},
  {Thayer}, {Thompson}, {Tibaldo}, {Tinivella}, {Torres}, {Tosti}, {Troja},
  {Usher}, {Vandenbroucke}, {Vasileiou}, {Vianello}, {Vitale}, {Winer}, {Wood},
  {Yamazaki}, {Younes}, {Yu}, {Zhu}, {Bhat}, {Briggs}, {Byrne}, {Foley},
  {Goldstein}, {Jenke}, {Kippen}, {Kouveliotou}, {McBreen}, {Meegan},
  {Paciesas}, {Preece}, {Rau}, {Tierney}, {van der Horst}, {von Kienlin},
  {Wilson-Hodge}, {Xiong}, {Cusumano}, {La Parola}, \&
  {Cummings}}]{2014Sci...343...42A}
{Ackermann}, M., {Ajello}, M., {Asano}, K., {et~al.} 2014, Science, 343, 42

\bibitem[{{Akaike}(1974)}]{1974ITAC...19..716A}
{Akaike}, H. 1974, IEEE Transactions on Automatic Control, 19, 716

\bibitem[{{Becerra} {et~al.}(2016){Becerra}, {Bianco}, {Fryer}, {Rueda}, \&
  {Ruffini}}]{2016ApJ...833..107B}
{Becerra}, L., {Bianco}, C.~L., {Fryer}, C.~L., {Rueda}, J.~A., \& {Ruffini},
  R. 2016, \apj, 833, 107

\bibitem[{{Becerra} {et~al.}(2015){Becerra}, {Cipolletta}, {Fryer}, {Rueda}, \&
  {Ruffini}}]{2015ApJ...812..100B}
{Becerra}, L., {Cipolletta}, F., {Fryer}, C.~L., {Rueda}, J.~A., \& {Ruffini},
  R. 2015, \apj, 812, 100

\bibitem[{{Becerra} {et~al.}(2018){Becerra}, {Ellinger}, {Fryer}, {Rueda}, \&
  {Ruffini}}]{2018ARep...62..840B}
{Becerra}, L., {Ellinger}, C., {Fryer}, C., {Rueda}, J.~A., \& {Ruffini}, R.
  2018, Astronomy Reports, 62, 840

\bibitem[{{Becerra} {et~al.}(2019){Becerra}, {Ellinger}, {Fryer}, {Rueda}, \&
  {Ruffini}}]{2019ApJ...871...14B}
{Becerra}, L., {Ellinger}, C.~L., {Fryer}, C.~L., {Rueda}, J.~A., \& {Ruffini},
  R. 2019, \apj, 871, 14

\bibitem[{Cash(1979)}]{1979ApJ...228..939C}
Cash, W. 1979, Astrophysical Journal, 228, 939

\bibitem[{{Damour} \& {Ruffini}(1975)}]{1975PhRvL..35..463D}
{Damour}, T., \& {Ruffini}, R. 1975, Physical Review Letters, 35, 463

\bibitem[{{Fryer} {et~al.}(2015){Fryer}, {Oliveira}, {Rueda}, \&
  {Ruffini}}]{2015PhRvL.115w1102F}
{Fryer}, C.~L., {Oliveira}, F.~G., {Rueda}, J.~A., \& {Ruffini}, R. 2015,
  Physical Review Letters, 115, 231102

\bibitem[{{Fryer} {et~al.}(2014){Fryer}, {Rueda}, \&
  {Ruffini}}]{2014ApJ...793L..36F}
{Fryer}, C.~L., {Rueda}, J.~A., \& {Ruffini}, R. 2014, \apjl, 793, L36

\bibitem[{{Heisenberg}(1924)}]{1924AnP...379..577H}
{Heisenberg}, W. 1924, NACA. Washington June 1951 (NACA TM 1291), Technical
  Memorandum , Translated from Annalen der Physik by National Advisory
  Committee For Aeronautics, 379, 577

\bibitem[{{J. Selsing} {et~al.}(2019){J. Selsing}, Fynbo, Heintz, Watson, , \&
  Dyrbye}]{GCN23695}
{J. Selsing}, Fynbo, J., Heintz, K., {et~al.} 2019, GRB Coordinates Network

\bibitem[{{J.D. Gropp} {et~al.}(2019){J.D. Gropp}, Kennea, Krimm, LaPorte,
  Lien, Moss, D.~M.~Palmer, \& Siegel}]{GCN23688}
{J.D. Gropp}, Kennea, J.~A., Krimm, N. J. K. P. H.~A., {et~al.} 2019, GRB
  Coordinates Network

\bibitem[{Landau(1937{\natexlab{a}})}]{Landau:480039}
Landau, L.~D. 1937{\natexlab{a}}, Collected papers of L.D. Landau, pp.193-216,
  Phys. Z. Sowjet., 11, 26.
\newblock \url{http://cds.cern.ch/record/480039}

\bibitem[{Landau(1937{\natexlab{b}})}]{Landau:480041}
---. 1937{\natexlab{b}}, Collected papers of L.D. Landau, pp.193-216, Phys. Z.
  Sowjet., 11, 545.
\newblock \url{http://cds.cern.ch/record/480041}

\bibitem[{Li(2018)}]{2018arXiv181003129L}
Li, L. 2018, arXiv.org, arXiv:1810.03129

\bibitem[{Meegan {et~al.}(2009)Meegan, Lichti, Bhat, Bissaldi, Briggs,
  Connaughton, Diehl, Fishman, Greiner, Hoover, van~der Horst, von Kienlin,
  Kippen, Kouveliotou, McBreen, Paciesas, Preece, Steinle, Wallace, Wilson, \&
  Wilson-Hodge}]{2009ApJ...702..791M}
Meegan, C., Lichti, G., Bhat, P.~N., {et~al.} 2009, The Astrophysical Journal,
  702, 791

\bibitem[{Meszaros \& Rees(2001)}]{Meszaros:2001vr}
Meszaros, P., \& Rees, M.~J. 2001, Astrophys. J., 556, L37

\bibitem[{{Piran}(2004)}]{2004RvMP...76.1143P}
{Piran}, T. 2004, Reviews of Modern Physics, 76, 1143

\bibitem[{{R. Hamburg} {et~al.}(2019){R. Hamburg}, Veres, Meegan, Burns,
  Connaughton, Kocevski, \& Roberts}]{GCN23707}
{R. Hamburg}, Veres, P., Meegan, C., {et~al.} 2019, GRB Coordinates Network

\bibitem[{{R. Mirzoyan} {et~al.}(2019){R. Mirzoyan}, Noda, Moretti, Berti,
  Nigro, Hoang, Micanovic, Takahashi, Chai, Moralejo, \& the
  MAGIC~Collaboration}]{GCN23701}
{R. Mirzoyan}, Noda, K., Moretti, E., {et~al.} 2019, GRB Coordinates Network

\bibitem[{{Robert M.~ Wald}(1974)}]{Wald:1974np}
{Robert M.~ Wald}. 1974, Phys. Rev., D10, 1680

\bibitem[{{Rueda} {et~al.}(2018){Rueda}, {Ruffini}, {Becerra}, \&
  {Fryer}}]{2018IJMPA..3344031R}
{Rueda}, J.~A., {Ruffini}, R., {Becerra}, L.~M., \& {Fryer}, C.~L. 2018,
  International Journal of Modern Physics A, 33, 1844031

\bibitem[{{Ruffini} {et~al.}(2019{\natexlab{a}}){Ruffini}, {Melon Fuksman}, \&
  {Vereshchagin}}]{RuffiniFuksman2019}
{Ruffini}, R., {Melon Fuksman}, J.~D., \& {Vereshchagin}, G.~V.
  2019{\natexlab{a}}, to be submitted

\bibitem[{Ruffini {et~al.}(1999)Ruffini, Salmonson, Wilson, \&
  Xue}]{1999A&AS..138..511R}
Ruffini, R., Salmonson, J.~D., Wilson, J.~R., \& Xue, S.-S. 1999, Astronomy and
  Astrophysics Supplement, 138, 511

\bibitem[{{Ruffini} {et~al.}(1999){Ruffini}, {Salmonson}, {Wilson}, \&
  {Xue}}]{1999A&A...350..334R}
{Ruffini}, R., {Salmonson}, J.~D., {Wilson}, J.~R., \& {Xue}, S.-S. 1999, \aap,
  350, 334

\bibitem[{{Ruffini} {et~al.}(2000){Ruffini}, {Salmonson}, {Wilson}, \&
  {Xue}}]{2000A&A...359..855R}
---. 2000, \aap, 359, 855

\bibitem[{{Ruffini} {et~al.}(2010){Ruffini}, {Vereshchagin}, \&
  {Xue}}]{2010PhR...487....1R}
{Ruffini}, R., {Vereshchagin}, G., \& {Xue}, S. 2010, \physrep, 487, 1

\bibitem[{{Ruffini} {et~al.}(2003){Ruffini}, {Vitagliano}, \&
  {Xue}}]{2003PhLB..573...33R}
{Ruffini}, R., {Vitagliano}, L., \& {Xue}, S.-S. 2003, Physics Letters B, 573,
  33

\bibitem[{{Ruffini} {et~al.}(2015){Ruffini}, {Wang}, {Enderli}, {Muccino},
  {Kovacevic}, {Bianco}, {Penacchioni}, {Pisani}, \&
  {Rueda}}]{2015ApJ...798...10R}
{Ruffini}, R., {Wang}, Y., {Enderli}, M., {et~al.} 2015, \apj, 798, 10

\bibitem[{{Ruffini} {et~al.}(2016){Ruffini}, {Rueda}, {Muccino}, {Aimuratov},
  {Becerra}, {Bianco}, {Kovacevic}, {Moradi}, {Oliveira}, {Pisani}, \&
  {Wang}}]{2016ApJ...832..136R}
{Ruffini}, R., {Rueda}, J.~A., {Muccino}, M., {et~al.} 2016, \apj, 832, 136

\bibitem[{{Ruffini} {et~al.}(2018{\natexlab{a}}){Ruffini}, {Rodriguez},
  {Muccino}, {Rueda}, {Aimuratov}, {Barres de Almeida}, {Becerra}, {Bianco},
  {Cherubini}, {Filippi}, {Gizzi}, {Kovacevic}, {Moradi}, {Oliveira}, {Pisani},
  \& {Wang}}]{2018ApJ...859...30R}
{Ruffini}, R., {Rodriguez}, J., {Muccino}, M., {et~al.} 2018{\natexlab{a}},
  \apj, 859, 30

\bibitem[{{Ruffini} {et~al.}(2018{\natexlab{b}}){Ruffini}, {Wang}, {Aimuratov},
  {Barres de Almeida}, {Becerra}, {Bianco}, {Chen}, {Karlica}, {Kovacevic},
  {Li}, {Melon Fuksman}, {Moradi}, {Muccino}, {Penacchioni}, {Pisani},
  {Primorac}, {Rueda}, {Shakeri}, {Vereshchagin}, \&
  {Xue}}]{2018ApJ...852...53R}
{Ruffini}, R., {Wang}, Y., {Aimuratov}, Y., {et~al.} 2018{\natexlab{b}}, \apj,
  852, 53

\bibitem[{{Ruffini} {et~al.}(2018{\natexlab{c}}){Ruffini}, {Becerra}, {Bianco},
  {Chen}, {Karlica}, {Kova{\v c}evi{\'c}}, {Melon Fuksman}, {Moradi},
  {Muccino}, {Pisani}, {Primorac}, {Rueda}, {Vereshchagin}, {Wang}, \&
  {Xue}}]{2018ApJ...869..151R}
{Ruffini}, R., {Becerra}, L., {Bianco}, C.~L., {et~al.} 2018{\natexlab{c}},
  \apj, 869, 151

\bibitem[{{Ruffini} {et~al.}(2018{\natexlab{d}}){Ruffini}, {Moradi}, {Rueda},
  {Becerra}, {Bianco}, {Cherubini}, {Filippi}, {Chen}, {Karlica}, {Sahakyan},
  {Wang}, \& {Xue}}]{2018arXiv181200354R}
{Ruffini}, R., {Moradi}, R., {Rueda}, J.~A., {et~al.} 2018{\natexlab{d}}, arXiv
  e-prints, arXiv:1812.00354

\bibitem[{{Ruffini} {et~al.}(2018{\natexlab{e}}){Ruffini}, {Rueda}, {Moradi},
  {Wang}, {Xue}, {Becerra}, {Bianco}, {Chen}, {Cherubini}, {Filippi},
  {Karlica}, {Melon Fuksman}, {Primorac}, {Sahakyan}, \&
  {Vereshchagin}}]{2018arXiv181101839R}
{Ruffini}, R., {Rueda}, J.~A., {Moradi}, R., {et~al.} 2018{\natexlab{e}}, arXiv
  e-prints, arXiv:1811.01839

\bibitem[{{Ruffini} {et~al.}(2019{\natexlab{b}})}]{GCN23715}
{Ruffini}, R., {et~al.} 2019{\natexlab{b}}, GRB Coordinates Network, 23715

\bibitem[{Schwarz {et~al.}(1978)}]{schwarz1978estimating}
Schwarz, G., {et~al.} 1978, The annals of statistics, 6, 461

\bibitem[{Vianello {et~al.}(2015)Vianello, Lauer, Younk, Tibaldo, Burgess,
  Ayala, Harding, Hui, Omodei, \& Zhou}]{2015arXiv150708343V}
Vianello, G., Lauer, R.~J., Younk, P., {et~al.} 2015, arXiv.org,
  arXiv:1507.08343

\bibitem[{{Wang} {et~al.}(2019{\natexlab{a}}){Wang}, {Li}, {Moradi}, \&
  {Ruffini}}]{2019arXiv190107505W}
{Wang}, Y., {Li}, L., {Moradi}, R., \& {Ruffini}, R. 2019{\natexlab{a}}, arXiv
  e-prints, arXiv:1901.07505

\bibitem[{{Wang} {et~al.}(2019{\natexlab{b}}){Wang}, {Rueda}, {Ruffini},
  {Becerra}, {Bianco}, {Becerra}, {Li}, \& {Karlica}}]{2019ApJ...874...39W}
{Wang}, Y., {Rueda}, J.~A., {Ruffini}, R., {et~al.} 2019{\natexlab{b}}, \apj,
  874, 39

\bibitem[{Wilson(1983)}]{RevModPhys.55.583}
Wilson, K.~G. 1983, Rev. Mod. Phys., 55, 583.
\newblock \url{https://link.aps.org/doi/10.1103/RevModPhys.55.583}

\bibitem[{{Xu} {et~al.}(2013){Xu}, {de Ugarte Postigo}, {Leloudas},
  {Kr{\"u}hler}, {Cano}, {Hjorth}, {Malesani}, {Fynbo}, {Th{\"o}ne},
  {S{\'a}nchez-Ram{\'{\i}}rez}, {Schulze}, {Jakobsson}, {Kaper}, {Sollerman},
  {Watson}, {Cabrera-Lavers}, {Cao}, {Covino}, {Flores}, {Geier}, {Gorosabel},
  {Hu}, {Milvang-Jensen}, {Sparre}, {Xin}, {Zhang}, {Zheng}, \&
  {Zou}}]{2013ApJ...776...98X}
{Xu}, D., {de Ugarte Postigo}, A., {Leloudas}, G., {et~al.} 2013, \apj, 776, 98

\bibitem[{Yu {et~al.}(2018)Yu, Dereli-B{\'e}gu{\'e}, \&
  Ryde}]{2018arXiv181007313Y}
Yu, H.-F., Dereli-B{\'e}gu{\'e}, H., \& Ryde, F. 2018, arXiv.org,
  arXiv:1810.07313

\end{thebibliography}
\end{document}